\documentclass[traditabstract]{aa} 
\usepackage{graphicx}
\usepackage{txfonts}
\usepackage{amssymb}
\usepackage{amsmath}
\newcommand{\sbsc}[1]{\ensuremath{_{\textrm{#1}}}}
\usepackage{gensymb}
\usepackage{natbib,twoopt}
\bibpunct{(}{)}{;}{a}{}{,} 
\makeatletter
\nonstopmode
\newcommandtwoopt{\citeads}[3][][]{\href{http://adsabs.harvard.edu/abs/#3}%
{\def\hyper@linkstart##1##2{}%
\let\hyper@linkend\@empty\citealp[#1][#2]{#3}}}
\newcommandtwoopt{\citepads}[3][][]{\href{http://adsabs.harvard.edu/abs/#3}%
{\def\hyper@linkstart##1##2{}%
\let\hyper@linkend\@empty\citep[#1][#2]{#3}}}
\newcommandtwoopt{\citetads}[3][][]{\href{http://adsabs.harvard.edu/abs/#3}%
{\def\hyper@linkstart##1##2{}%
\let\hyper@linkend\@empty\citet[#1][#2]{#3}}}
\newcommandtwoopt{\citeyearads}[3][][]%
{\href{http://adsabs.harvard.edu/abs/#3}
{\def\hyper@linkstart##1##2{}%
\let\hyper@linkend\@empty\citeyear[#1][#2]{#3}}}
\makeatother
\usepackage[colorlinks=true, linkcolor=blue, citecolor=blue,urlcolor=blue]{hyperref}
\usepackage{caption}

\begin{document}
\title{How do starspots influence the transit timing variations of
exoplanets? Simulations of individual and consecutive transits.}
\author{P. Ioannidis, K.F. Huber \and J.H.M.M. Schmitt}
\institute{Hamburger Sternwarte, Universit\"at Hamburg, Gojenbergsweg 112,
21029 Hamburg, Germany\\
\email{pioannidis@hs.uni-hamburg.de}}
\date{Received 18 August 2015 / Accepted 09 October 2015}
\abstract{ Transit timing variations (TTVs) of exoplanets are 
normally interpreted as the consequence of gravitational interaction with 
additional  bodies in the system.
However, TTVs can also be caused by deformations of the system transits by starspots, 
which might thus pose a serious complication in their interpretation. 
We therefore simulate transit light curves deformed by spot-crossing events 
for different properties of the stellar surface and the planet, such as 
starspot position, limb darkening, planetary period, and impact parameter. 
Mid-transit times determined from these simulations can be significantly shifted 
with respect to the input values; these shifts cannot be larger than ~1\% of the 
transit duration and depend most strongly on the longitudinal position of the spot during 
the transit and the transit duration. 
Consequently, TTVs with amplitudes larger than the above limit are very unlikely to 
be caused by starspots. \\
We also investigate whether TTVs from sequences of consecutive 
transits with spot-crossing anomalies can be misinterpreted 
as the result of an additional body in the system. 
We use the Generalized Lomb-Scargle periodogram to search for periods in TTVs 
and conclude that low amplitude TTVs with statistically significant periods 
around active stars are the most problematic cases.
In those cases where the photometric precision is high enough to inspect the transit shapes 
for deformations, it should be possible to identify TTVs caused by starspots, 
however, especially for cases with low transit signal to noise light curves 
(TSNR~$\lesssim 15$) it becomes quite difficult 
to reliably decide whether these periods come from 
starspots, physical companions in the system or if they are random noise 
artifacts.}

\titlerunning {Spot-crossing events and TTVs}
\maketitle

\section{Introduction}

The transits of a planet in front of its host star contain a wealth of fundamental
information both on the planet and its host star. 
Using mid-transit times and studying their variations
it is possible to measure the gravitational interaction of additional bodies that orbit the same center of mass.
These transit timing variations (TTVs) can be used to determine the masses of planets
in a multi-planet system and even have the potential of inferring the existence
of additional (even low mass) companions as has been shown, for example, by \citetads{2005MNRAS.359..567A}.
Although masses and additional companions can also be inferred from precise radial velocity (RV) measurements,
the importance of the TTV method increases for planets around faint stars where accurate RV
measurements are difficult to obtain.

Depending on the actual system configuration, TTVs can reach differences 
w.r.t. the linear period ephemeris of up to a few hours,
especially for systems close to low order mean motion resonances \citepads{2013ApJ...777....3N}.
Although the determination of TTVs is possible from ground-based observations, 
the high precision of the data from the {\it CoRoT} \citepads{2006ESASP1306...33B} 
and {\it Kepler} \citepads{2010ApJ...713L..79K} space missions, along with the 
possibility for uninterrupted sequences of transit observations,
is better suited for this task and has provided evidence for statistically significant TTVs in about
$60$~cases.\footnote{according to the list provided by http://exoplanet.org as of May 2015} 

Space-based photometry has further shown that the light curve of a star can also be substantially 
disturbed by effects of stellar activity.
In particular, the transit profiles of planets with active host stars can be heavily distorted by dark spots as observed,
for example, in Tres-1~b \citepads{2009A&A...494..391R} or
\mbox{CoRoT-2~b} (e.g., \citeads{2009A&A...504..561W}, \citeads{2009A&A...508..901H}).
This effect can be severe enough to introduce large systematic errors
in the determination of the physical parameters of the planets,
which can by far exceed the formal statistical errors (\citeads{2009A&A...505.1277C}, \citeads{2013A&A...556A..19O}).

Recently, have been reported the detection of low amplitude TTVs
which are assumed to be caused by starspot signatures in the transit profiles
(\citeads{2011ApJ...733..127S}, \citeads{2012ApJ...750..114F}, \citeads{2013A&A...553A..17S},
\citeads{2013ApJS..208...16M}).
A first approach to study the effect of starspots on TTVs was therefore made
by \citetads{2013MNRAS.430.3032B} to explain the $\sim$3~min TTVs of the
planet WASP-10~b, while a more general approach,
including a variety of configurations between the starspot and the planet,
was presented by \citetads{2013A&A...556A..19O}.

In this paper we simulate light curves of exoplanetary transits including
spot-crossing events for a wider variety of the properties of the underlying star, starspot, and planet,
and investigate the influence of these starspot signatures on the determination of mid-transit times.
Furthermore, we study sequences of consecutive transits affected by starspots
and evaluate the possibility of detecting a statistically significant signal in the TTVs
that could mimic the TTV signal of a physical body.

The structure of our paper is as follows:
In Sect.~\ref{Sec:LCsims} we describe our approach of generating light curves in general,
including our basic assumptions on the properties of the stellar surface and the planet.
In Sect.~\ref{par:strat} we state the specific questions we want to investigate
and describe the individual simulation setups in detail.
In Sect.~\ref{par:res} we present the results for each of our previously described simulation setup.
In Sect.~\ref{par:sct} we explore the possibility for the spot-crossing events to create spurious TTVs in
transit sequences, and finally, in Sect.~\ref{Sec:Discussion}
we discuss our results, including a comparison with real TTV observations,
and conclude with a summary.

\section{Light curve simulations}
\label{Sec:LCsims}

\subsection{Light curves of spotted stars}

A rotating star with a persistent dark spot on its surface
produces a periodic modulation of its light curve.
However, the exact shape of the photometric variation depends both
on the properties of the star and the spot.
If the life time $\tau\sbsc{sp}$ of the spots (or rather active regions)
is much longer than the stellar rotation period $P_{\star}$,
the observed flux variations will be similar from one rotation to the
next.

In many of the {\it CoRoT} and {\it Kepler} light curves of active stars with
$\tau\sbsc{sp}/P_\mathrm{\star} > 1$
the observed photometric variations consist of one or two pronounced dips per rotation,
usually without the presence of any intervals where the stellar flux stays constant.
The latter suggests that there are always at least a few spots located on
the star that rotate on and off the visible hemisphere.
The peak-to-peak amplitude of such photometric variations
can be as high as several percent of the total flux \citepads{2013ApJ...769...37B}.

Throughout our simulations we make the simplified assumption that
the stellar surface is covered by only one (large) spot
with a temperature lower than the photospheric temperature.
We further assume that the spot properties do not change, i.e.,
we do not consider spot evolution effects or
any internal spot structure such as penumbra and umbra.
We adopt a center-to-limb intensity variation of the stellar photosphere
according to the specific temperature
which is parameterized using a quadratic limb-darkening law.

\subsection{Spot-crossing events}\label{par:scev}

In the following we briefly explain our mathematical description
of the photometric effects we expect during a spot-crossing event.
During a planetary transit the total flux $F_0$ of the star decreases to $F_\mathrm{tr}$
because the planet covers part of the stellar disk. We obtain for the ratio $F_\mathrm{tr}/F_0$
the expression
\begin{equation}
\label{eq:planet}
\frac{F_\mathrm{tr}}{F_0} \, = \, 1 \, - \, \frac{R\sbsc{pl}^2}{R_{\star}^2} \, = \, 1 \, - \, f_\mathrm{pl},
\end{equation}
where $R\sbsc{pl}$ denotes the planetary radius and $R_{\star}$ the stellar radius. 
We call $f_\mathrm{pl}$ the planetary filling factor.
For the sake of simplicity, we assume that the planetary disk is completely dark,
we ignore the limb-darkening (LD) effects
and the ingress and egress phases, when the stellar disk is only partially
covered by the planet; however, the formalism is easily extended to cover these effects.

Let us now consider a spotted star with a spot filling factor
$f\sbsc{sp}$, which denotes the relative area covered by spots. 
We also introduce the relative spot temperature
\mbox{$t\sbsc{sp}=T\sbsc{sp}/T_{\star}$},
where $T\sbsc{sp}$ and $T_{\star}$ denote the temperatures of the spot and the photosphere, respectively.
In this case the observed normalized flux of the star 
is not $F_0$, i.e., the flux one would
see from an undisturbed photosphere.
Rather we see the ''spotted'' flux $F_\mathrm{sp}$ given by the expression
\begin{equation}
\label{eq:fobs}
\frac{F_\mathrm{sp}}{F_0} \, = \, 1 \, - \, f\sbsc{sp} ( 1 - t\sbsc{sp}^4 ) \ .
\end{equation}

Now consider the case of a spot-crossing event during a planetary transit and assume that
the transiting planet is occulting a starspot either partially or totally.
Let $df\sbsc{pl}$ denote that fraction of the planetary disk covering spotted surface area,
while ($1 - df\sbsc{pl}$) denotes the fraction of the planetary disk covering unspotted photosphere.
In this case, the observed flux during the transit
$F_{tr,sp}$ is given by the ''spotted'' flux $F_\mathrm{sp}$,
diminished by the relative flux of spotted and unspotted photosphere
covered by the planet, i.e.,

\begin{figure}[bp]
\includegraphics[width=\linewidth]{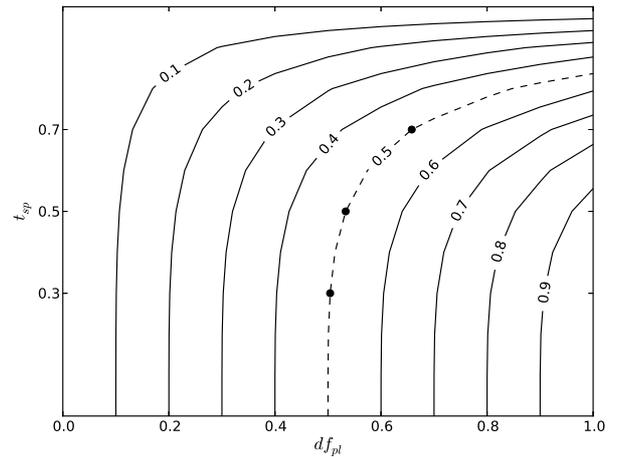}
\caption{The spot-crossing anomaly height $H$ (contours)
as a function of the fraction of the planetary disk 
which covers spotted surface area $df\sbsc{pl}$ and the spot 
temperature $t\sbsc{sp}$.
For our simulations we use values of $df\sbsc{pl}$ that correspond
to spot temperatures equal to 0.3, 0.5, and 0.7 in order to reach $H=0.5$ 
(black dots along the dashed line).}
\label{mtrflx}
\end{figure}

\begin{equation}
\label{eq:combined}
\frac{F_\mathrm{tr,sp}}{F_0} \, = \, 1 - f\sbsc{sp} ( 1 - t\sbsc{sp}^4 ) \, - \, f_\mathrm{pl}
\left[ 1 - df_\mathrm{pl} \left( 1 - t\sbsc{sp}^4 \right) \right].
\end{equation}

We usually work with transits normalized to the out-of-transit flux,
which means we divide this equation by the contribution of the spot (Eq.\ref{eq:fobs}.
This normalized flux $d$ is then defined by

\begin{equation}
\label{eq:norm-oot}
d \, = \frac{F_\mathrm{tr,sp}}{F_{sp}} = 
\, 1 \, - \, \frac{f_\mathrm{pl} \left[ 1 - df_\mathrm{pl} \left( 1 -
t\sbsc{sp}^4 \right) \right]}{1 - f\sbsc{sp} ( 1 - t\sbsc{sp}^4 )} \ .
\end{equation}

If the starspot only insignificantly contributes flux with respect to the unspotted surface
($t\sbsc{sp} \approx 0 $), and if that spot is large enough that the planet
covers only spotted surface regions ($df_\mathrm{pl} = 1$),
Eq.~\ref{eq:norm-oot} becomes unity, i.e.,
the spot-crossing event would increase the flux to the out-of-transit level.
We are not aware of any observation presented so far
where this has been seen and we propose several reasons for this:
First, although starspots are likely to be much cooler than photospheric
temperatures, they still do contribute a significant amount of flux.
Second, most individual starspots are probably not as large as a planet.
When we talk about starspots, we usually do not think of
one large uniformly dark spot but of a densely spotted region made up of umbra,
penumbra, and photospheric parts.
However, such an active region might easily be much larger than the occulting
planet, although the very dark (umbral) parts of the region are not.
Third, a planet crossing over such a spotted area does not always have to cover the
darkest parts of the spot but can cross near the periphery as well.

We introduce the height $H$ of the spot-crossing anomaly in a transit,
which is given by the comparison between the relative flux $d$ in the cases
with and without ($d_{df\sbsc{pl}=0}$) a spot-crossing event.
\begin{equation}
\label{eq:bumpsize}
H \, = \, \frac{d \, - \, d_{df\sbsc{pl}=0}}{1\, - \, d_{df\sbsc{pl}=0}} \, = \, 
df\sbsc{pl} ( 1 - t\sbsc{sp}^4 ).
\end{equation}
Figure~\ref{mtrflx} illustrates the dependence of $H$ on
$df\sbsc{pl}$ and $t\sbsc{sp}$;
for \mbox{$H=0$} there is no spot-crossing anomaly, while $H=1$ means that the
increase is equal to the transit depth and, thus,
the flux is back to out-of-transit level.
As discussed before, values of $H > 0.5$ have not been observed so far
and seem to be unlikely; therefore, we select those combinations of $df\sbsc{pl}$ and $t\sbsc{sp}$
in our simulations that produce an increase in flux of $50$~\% of the mid-transit depth
(see Fig.~\ref{mtrflx}, dashed line).

\subsection{Model calculations}

Given the assumption that a spot has a either circular or an elliptical shape
with a total area $A_\mathrm{sp}$,
our algorithm calculates analytically the light curve modulations
due to stellar rotation using the relationship
\begin{equation}
\label{eq:spotflux}
\frac{F_\mathrm{sp}}{F_0} \, = \, 1 \, - \, \frac{A_\mathrm{sp} \cdot \cos(\theta)}{\pi R_{\star}^{2}} \, ( 1 - t\sbsc{sp}^4 )
\cdot I(\theta,c_1,c_2) \ ,
\end{equation}
where $\theta$ denotes the angle between the normal to the stellar
surface and the line of sight towards the observer.
The factor $I(\theta,c_1,c_2)$ accounts for the LD
of the disk which depends on the angle $\theta$
and, because we choose to use a quadratic LD parameterization,
on the limb-darkening coefficients (LDC) $c_1$ and $c_2$.

\begin{figure}[tp]
\includegraphics[width=\linewidth]{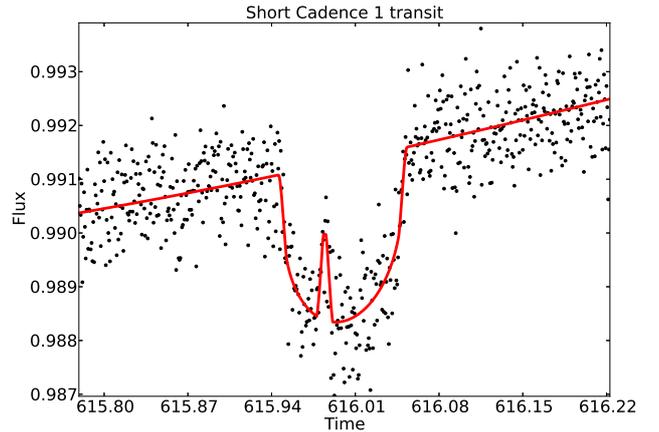}
\caption{Simulated transit of a Neptune-sized planet.
The red line represents the infinite signal-to-noise data, while 
the noise of the black dots is similar to that of 
{\it Kepler} short cadence data.
During the transit there is a spot-crossing event right after the ingress.}
\label{spcrtr}
\end{figure}

When a planet transits in front of the star,
our algorithm uses the model of \citetads{2002ApJ...580L.171M} to calculate the
transit-induced flux reduction.
In Fig.~\ref{spcrtr} we show an example in order to demonstrate some typical light
curve that our algorithm produces.
In this example the light curve of a Neptune-sized planet orbiting a K-type dwarf is shown;
its noise level is adjusted to a star with a {\it Kepler} magnitude of
\mbox{$K_p = 13$} (black dots) along with the infinite signal to noise case (red line).

The calculation of spot-crossing events involves intersections
of circles and an ellipse, which we choose to perform numerically.
To the left of the transit center there is an anomaly caused by a spot-crossing event.
The overall light curve slope is caused by the flux variation due to the rotation of the star.

\subsection{General assumptions}
\label{par:assum}

We briefly summarize the basic assumptions for the simulated light curves
we use in order to study the effects of spot-crossing events on TTVs:

\begin{enumerate}
 \item The planet crosses only one uniformly dark spot per transit.
 \item The shape of the spot is either circular or elliptical.
 \item We consider relative spot temperatures $t\sbsc{sp}$ of 0.3, 0.5, and 0.7.
 \item The fraction of the planet that covers spotted area is chosen in such a way that
       the flux increment at mid-transit yields $50$~\% \mbox{($H=0.5$)}, 
       assuming an impact parameter $b=0$.
 \item The noise level is adjusted to the precision of {\it Kepler} short cadence
       photometry for a $K_p=13~\mathrm{mag}$ star \citep{2010ApJ...713L.160G}.
 \item We introduce the parameter signal-to-noise ratio of the transit (TSNR)
       which is the transit depth divided by the out-of-transit light curve noise level (see assumption $5$).  
 \item Each data point represents an integration over $60$~seconds.
 \item The rotation period of the star is fixed to $10$~days.
\end{enumerate}

\section{Simulation strategy}
\label{par:strat}

Spot-crossing events cause deformations of the transit shape
which, potentially, lead to a shift of the measured mid-transit time.
We set up individual light curve simulations to address the 
following three questions:

\begin{enumerate}
 \item How much are the measured mid-transit times affected by the (longitudinal)
       position of a spot during a spot-crossing event?
 \item How do the orbital characteristics of the planet, i.e, its orbital period and impact parameter,
       influence the shift of mid-transit times?
 \item Does the influence of a spot-crossing event on timing measurements vary for different stellar types?
\end{enumerate}

\subsection{Spot position}
\label{Sec:SS-spotpos}

The strength of the deformation of a transit profile caused by
a spot-crossing event depends on the longitude of the spot.
To study the effect on mid-transit times,
we first consider a transit of a planet with an orbital period equal to the
stellar rotation period.
The planet has an impact parameter of \mbox{$b=0$} and,
thus, transits the center of the stellar disk.
A spot with a relative temperature of \mbox{$t\sbsc{sp} = 0.3$} is placed on the stellar equator
at longitude~$\lambda\sbsc{o}$.
$\lambda\sbsc{o}$ is the longitude of the spot at the epoch of the observation;
$\lambda\sbsc{o}=0$ denotes the sub-observer point at the center of the disk,
with $\lambda\sbsc{o}=-90$ being the leading edge and $\lambda\sbsc{o}=90$ the
trailing edge of the disk.

We adopt a K-type dwarf host star \mbox{($T_{eff} = 4300~K$)} with a quadratic
LD law and use LDCs of $c_1 = 0.690$ and $c_2 = 0.034$
as provided by \citetads{2012yCat..35469014C} for the $\it Kepler$ mission.
We simulate light curves based on three different assumptions for TSNR:
one for infinite TSNR, and two more with TSNR values equal to the expected precision for planets with
$R\sbsc{pl}/R_{\star}=0.05$ and $R\sbsc{pl}/R_{\star}=0.1$ (see assumptions $5$~\&~$6$
in Sect.~\ref{par:assum}).

\begin{figure}[tp]
\includegraphics[width=\linewidth]{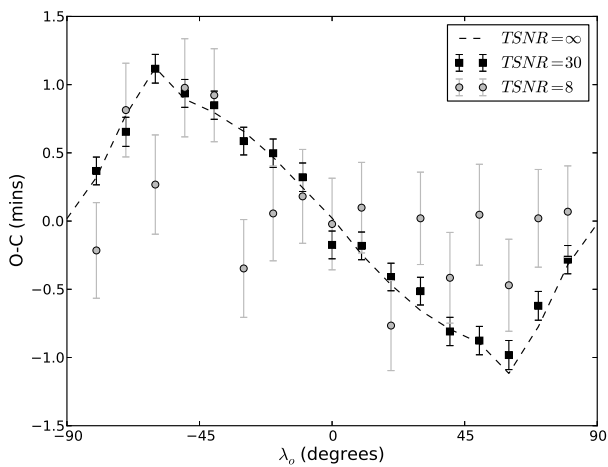}
\captionof{figure}{
The $t\sbsc{sp}=0.3$ simulation 
for different TSNR levels.
The dashed line represents the case of infinite TSNR.
The measurement error for transits with TSNR=30
is still small enough to see the underlying modulation caused by the spots,
whereas for TSNR=8 the spot signal vanishes in the noise.
}
\label{infvssnr}
\end{figure}

\subsection{Planetary orbit}
\label{par:porb}

If the planet's orbital period $P\sbsc{pl}$ is much smaller
than the stellar rotation period $P_{\star}$,
the spot's position hardly changes while the planet is crossing over it.
If $P\sbsc{pl}/P_{\star}$ increases, the spot starts to move significantly
during its occultation by the planet.
To estimate the strength of this effect on the measurement of mid-transit times,
we study simulations with orbital periods of 5, 10, 20, 40, 80, 150, 300 and 600~days.
We also consider different impact parameters of $0\leq b<1$ in combination with different spot longitudes $\lambda\sbsc{o}$.
The sizes of our test planets are $R\sbsc{pl}/R_{\star}=0.1$ and $R\sbsc{pl}/R_{\star}=0.05$.
For this simulation the stellar inclination is fixed to zero and the host star
has the same properties as in the Sect.~\ref{Sec:SS-spotpos}.

\subsection{Stellar limb darkening and inclination}
\label{Sec:SS-starchar}

Even though the effect of a spot-crossing event on TTV measurements should be
dominated by the properties of the spot and the planet,
we also investigate the influence of LDCs in detail.
We consider main sequence stars with effective temperatures of $3500~\mathrm{K}
\leq T\sbsc{eff} \leq 10\,000$~K and use the corresponding LDCs of the
{\it Kepler} band pass \citepads{2012yCat..35469014C}.
Furthermore, we investigate how different stellar inclinations influence the measurement of mid-transit times.
The planet in this setup has a size of $R\sbsc{pl}/R_{\star}=0.1$ and an impact parameter of $b=0$. 
We vary the inclination of the star as well as the spot latitude.

\section{Results}
\label{par:res}

The light curves simulated in Sect.~\ref{par:strat}
generally contain a slope caused by the starspot
located on the rotating surface (see Fig.~\ref{spcrtr}).
In order to remove this trend, we first identify the transit in the light curve,
remove those parts of the light curve affected by the transit,
fit a second order polynomial to the remaining out-of-transit data,
and divide the fit from out-of and in-transit data to produce a normalized transit light curve.
We then compute the mid-transit times and errors from these normalized transit light curves by
using the model described by \citeads{2002ApJ...580L.171M}
combined with a Markov-Chain Monte-Carlo (MCMC) sampler, 
with the only free parameter being the mid-transit time.

\subsection{Spot position}
\label{Sec:RES-spotpos}

Here we present the results derived for the simulations described in
Sect.~\ref{Sec:SS-spotpos}.
In Fig.~\ref{infvssnr} (dashed line) we show the difference (\mbox{$O-C$}) 
between mid-transit times
determined from our simulated light curve and the input times
for the different longitudes $\lambda\sbsc{o}$ of the occulted spot.

Figure~\ref{infvssnr} (dashed line) also illustrates that the maximum 
deviation of TTVs measurements
caused by spot-crossing events is $\sim 1.2$~minutes in this particular case.
The maximum values are encountered when the spot is located between the
center and the limb of the stellar disk at about $\lambda\sbsc{o}\sim \pm 70\degree$.
An anomaly at the transit center has no influence on the derived mid-transit times.
The spot-induced deviations also decrease towards the limb of the star
because the projected size of the spot decreases
and LD reduces the flux contributing from these parts of the stellar disk.

The spot-crossing anomalies introduce positive variations when they are
occurring before the mid-transit time and negative ones in the opposite case.
When the planet occults a spot during ingress or egress,
the measured transit duration is affected mostly due to the fact
that there is a pseudo-shift of the first or forth contact time;
the transit seems to start later or end earlier, respectively.
Thus, there is a shift of the mid-transit time into the opposite direction of the
spot-crossing event.

So far we have dealt with the case of infinite TSNR.
Naturally, measured mid-transit times have errors
which we determine using a MCMC algorithm.
Measurements and their errors
for the case of a planet with $R\sbsc{pl}/R_{\star}=0.1$
are shown in Fig.~\ref{infvssnr}, which corresponds to an TSNR of 30,
and $R\sbsc{pl}/R_{\star}=0.05$ equivalent to \mbox{TSNR=8}.
For the larger planet the spot-induced deviations of the mid-transit times from the input
is still detectable within the relatively small errors
and follows the same pattern as for the infinite TSNR case.
However, the errors for the smaller planet are much higher
and a systematic modulation caused by spots is no longer clearly visible.

\begin{figure}[tp]
\includegraphics[width=\linewidth]{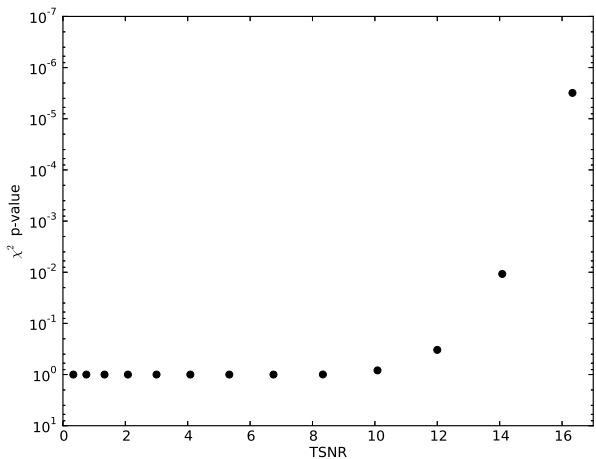}
\caption{The p-values of the null hypothesis $\chi^2$ analysis for mid-transit
times of transits with different TSNR values. The transits with TSNR < 15 do not show
statistically significant evidence of variations. As a result the 
mid-transit times of better resolved
light curves are more vulnerable to deviations caused by spot crossing events.}
\label{snrana}
\end{figure}

Assuming as null hypothesis that no TTVs are produced by the spot, 
we perform a $\chi^2$-analysis on the TTVs (cf. Fig~\ref{infvssnr}) of 
transits with different TSNR.
As shown in Fig.~\ref{snrana}, the deviations
caused by spot-crossing events start to disappear in the measurement noise for
transits with TSNR$\lesssim$15.

\begin{figure}[bp]
\includegraphics[width=\linewidth]{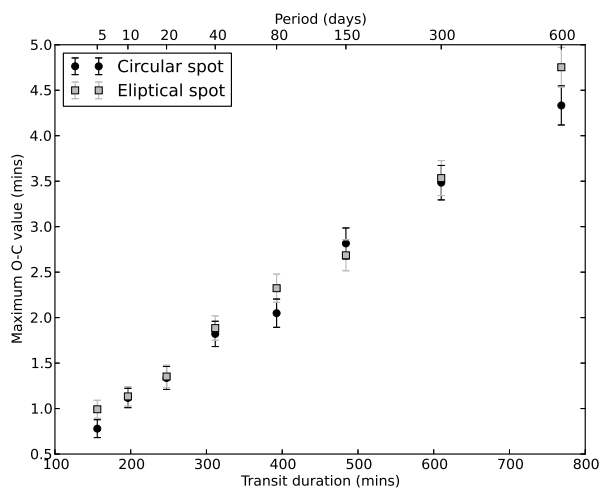}
\caption{The maximum deviation in the mid-transit time of a planet
with TSNR=30 for different transit durations.
As the period and hence the transit duration increases, the maximum deviation grows larger.
The black dots are referring to the maximum values caused by circular spots;
the gray squares refer to elliptical spots.}
\label{cirvsel}
\end{figure}

\begin{figure}[bp]
\includegraphics[width=\linewidth]{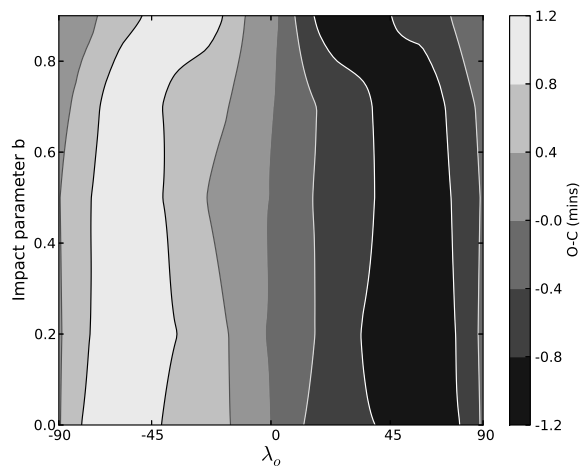}
\caption{TTVs for a simulation of a planet with TSNR=30 (contour map)
as a function of the longitude $\lambda_{o}$ and the impact parameter b.
As shown in Fig.~\ref{infvssnr}, $O-C$ has positive values for spot-crossing events before $\lambda\sbsc{o} = 0$ and negative values otherwise. While there is hardly a difference in the maximum values,
their positions shift to longitudes closer to the center of the star for higher impact parameters.}
\label{bvar}
\end{figure}

\subsection{Planetary orbit}
\label{sc:porb}

Based on the simulations described in Sect.~\ref{par:porb},
Fig.~\ref{cirvsel} presents the maximum deviation of mid-transit time
measurements caused by spot-crossing events for different values of $P\sbsc{pl}/P_{\star}$.
The TTV maximum increases linearly with the transit duration, which in turn
increases with the period ratio. For the same spot-crossing event a 
longer transit duration leads to a larger shift (e.g., \citeads{2011ApJ...733..127S}).

We also consider the possibility that spots might not always be circular.
In Fig.~\ref{cirvsel} we additionally show our results for elliptical spots,
which indicate that circular and elliptical spots with equal filling factors
lead to virtually identical results.
We conclude that the precise spot geometry appears to be of minor importance.

Next we consider how the derived transit times depend on
the impact parameter of the transits.
Figure~\ref{bvar} shows how the difference between measured and calculated
mid-transit times changes with the impact parameter of the planet. 
Due to the LD of the stellar disk and geometrical effects,
the transit profile is becoming more V-shaped for larger impact parameters, and
the maximum of the residuals moves closer to the center of the star.
The change of the transit shape is increasing the influence of 
the spot-crossing event on the measurement of the mid-transit time. 
On the other hand, for high impact parameters, the amplitude of the $O-C$ values 
becomes smaller due to the shorter transit duration (cf. Fig.~\ref{cirvsel}). 
Therefore, the additional deviation that the 
deformation of the transit introduces is canceled by the reduction of 
the transit duration. As a result, there is no significant 
difference in the amplitude of the residuals for different impact parameters.

\begin{figure}[tp]
\includegraphics[width=\linewidth]{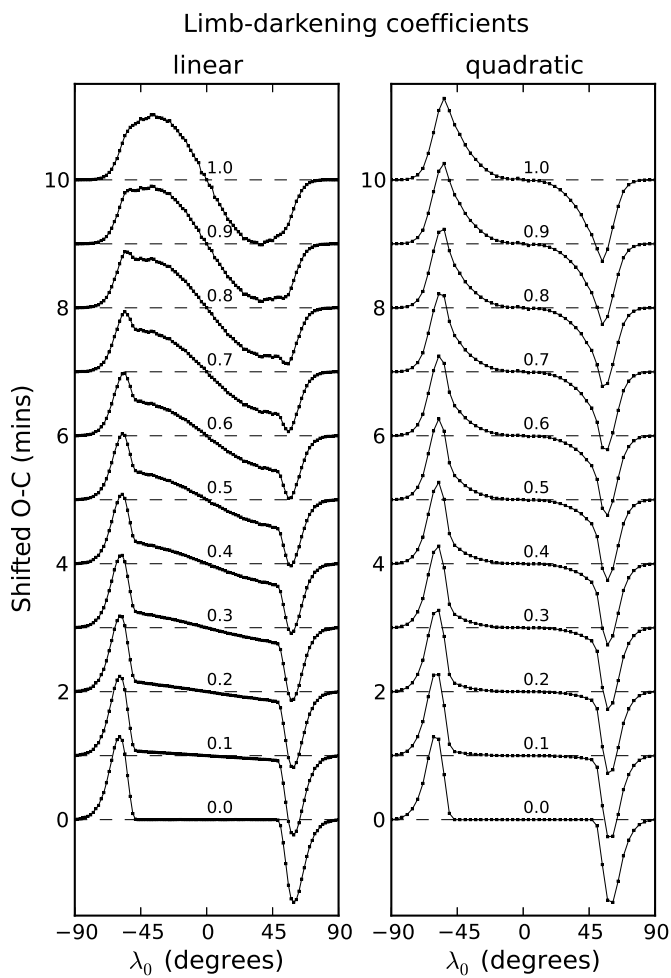}
\caption{The effect of different LDCs on the measured mid-transit time
for transits with infinite TSNR and spot-crossing events in various
longitudes $\lambda\sbsc{o}$.
The panels refer to linear (left) and quadratic (right) LDCs with values between unity (top) and zero (bottom). }
\label{ldldc}
\end{figure}

\subsection{Stellar limb darkening and inclination} 
\label{Sec:res-starchar}

In order to study how LD affects the we used a planet size of $R\sbsc{pl}/R_{\star}=0.1$ and a
relative spot temperature of $t\sbsc{sp}=0.3$, assuming a period ratio equal to unity.
The strength of the measured deviations does not significantly depend on the LDCs.
However, for lower effective temperatures the region of large deviations introduced to the mid-transit time
is larger and its inner boundary moves closer to the center of the star.

To better illustrate the influence of LD on the effect of transit time measurements,
we simulated light curves with different linear and quadratic LDCs in the range of zero to unity.
The results are presented in Fig.~\ref{ldldc},
where the left panel show the results for different linear LDCs with $c_2=0$
and for different quadratic LDCs with $c_1=0$ (right panel).
While the spot-crossing anomaly is moving from one side of the transit to the other,
the measured shift is changing similar to what has been shown in, e.g., Fig.~\ref{infvssnr}.
However, Fig.~\ref{ldldc} illustrates that the
influence of LD is rather large on the shape of the curve.
For small LDC values there is only a spot-induced shift of the transit time at ingress and egress
and hardly a shift in between.
This results from the fact that in the absence of LD the transit has
a box-shaped profile and is flat at the bottom.
The model does not have the flexibility to compensate an anomaly
caused by the spot in the flat part of the transit.
Thus, this anomaly cannot be fitted and the best fit results ignore the spot.

However, this changes for larger LDCs
because the transit shape becomes rounder
and the bottom is not flat anymore.
The stronger the LD effect the higher the curvature,
and a shift of the model can lead to a better fit
even if the spot signature is closer to the center.
This is especially true for the linear component
because it influences the center of the transit
much more than the quadratic LDC.
The latter has its largest influence at the ingress and
egress which is the reason why in this case the curves
change most in the outer parts and only little at the center.

Although the influence of LD appears to be large (see Fig.~\ref{ldldc}),
for the relevant temperatures considered in our
simulations the LDCs change only little and 
stay around roughly \mbox{$c_1=0.7$} and \mbox{$c_2=0.1$}.
For small changes in the LDCs the curves in 
Fig.~\ref{ldldc} change only slightly.

Finally, we also investigated how different values for the inclination
of the rotation axis of the star affect the measured transit times.
However, we could not detect a dependence in the maximum amplitude of the 
spot-induced TTVs on the stellar inclination.

\section{Sequence of consecutive transits}
\label{par:sct}

So far we studied only the effects of a spot-crossing event on the
TTV measurements of a single transit for different parameter combinations.
However, we have not yet investigated series of consecutive transits of the same system.
According to our previous results, a spot-crossing event can cause
deviations of the mid-transit times of up to a few minutes.
The amplitude alone, however, does not tell us whether it is possible
to produce a spurious signal in the timing measurements from a sequence of transits
that could mimic the signal of a true physical companion in the system.
In this case the activity would lead to a false positive detection of an
actually non-existent body.

It is rather difficult to study such problems in simulations
because the results always sensitively depend on the selected input parameters.
If we choose some specific spot configuration
with spot lifetimes and differential rotation, etc.,
our results would strongly depend on these assumptions.
We try to minimize this by choosing a simple, straight-forward
approach which considers a worst-case scenario:
We assume that for all transits in a light curve
the spot is always occulted if it is located on the visible hemisphere 
and that the spot has an infinite lifetime.
This is certainly not what one expects to find in reality;
spots are not always located exactly under the planet path
and they certainly have limited lifetimes.
Furthermore, we adopt a comparatively large spot.
However, in this way our results will give an estimate on what might be
observed in the most problematic cases.

\subsection{Simulation setup}

\begin{figure}[tp]
\includegraphics[width=\linewidth]{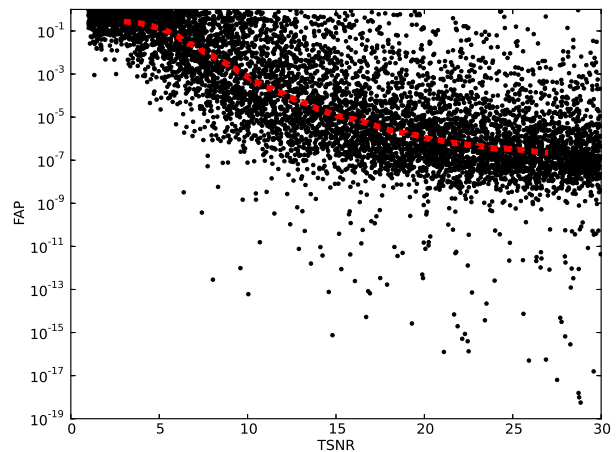}
\caption{The FAP of the highest peak in the Generalized Lomb-Scargle periodogram
over the TSNR of the planetary transits for all 7500~simulation.
On average, the strongest periods in the simulations become more significant for higher TSNRs.
The red line demonstrates the running median of the FAP in TSNR bins of six.
The majority of points larger than TSNR~$\gtrsim 15$ have FAPs below $10^{-3}$.}
\label{fapsnr}
\end{figure}

We set up 7500 simulations in such way that for each individual simulation
a series of 100 consecutive transit light curves of the same system are calculated.
We consider different TSNR values and orbital periods of the planet (measured in days),
which are both randomly drawn from a uniform distribution between 1 and 30.
The stellar rotation period is always fixed to 10~days.
We assume only one spot on the surface which is always occulted
when located on the visible hemisphere.
If the planet crosses an unspotted disk, its transit 
will not be deformed by any spot-crossing feature;
nonetheless, the transit will continue to be part of the sequence.
The spot lifetime is assumed to be infinite and its covered area is set to follow the assumption given 
in item~4 of Sect.~\ref{par:assum}.

We then measure the mid-transit times of the simulated transits as described in Sect.~\ref{par:res}
(including their measurement uncertainties using MCMC)
and calculate the residuals by subtraction the expected values ($O-C$).
Additionally, we calculate the mean of the residuals and subtract it.
The Generalized Lomb-Scargle periodogram \citepads{2009A&A...496..577Z} is then used to
determine the period $P_\mathrm{max}$ in the TTVs with the smallest false alarm probability (FAP)
for each individual simulation.
Finally, for each simulation we calculate the amplitude of the TTVs using the relationship,
\begin{equation}
A_\mathrm{max} = [(O-C)_\mathrm{max}-(O-C)_\mathrm{min}]/2.
\label{amax} 
\end{equation}

\subsection{Simulation results}
\label{Sec:SCT-res}

In Fig.~\ref{fapsnr} we show the FAP of the most significant period as a function of the TSNR
for each of our simulations.
The average FAP decreases with increasing TSNR.
This reflects the fact that transit times calculated from light curves with small TSNRs
have relatively large uncertainties which leads to higher FAPs when searching for periods.
The majority of periods for transits with TSNRs~$\gtrsim 15$ have FAPs~$\lesssim 10^{-3}$,
which we considered significant enough not to be spurious signals.
Although in our simulations these periods are caused by spots,
a reliable distinction whether such a period is caused by spots or additional bodies in the system
might be quite complicated in real observations.

\begin{figure}[tp]
\includegraphics[width=\linewidth]{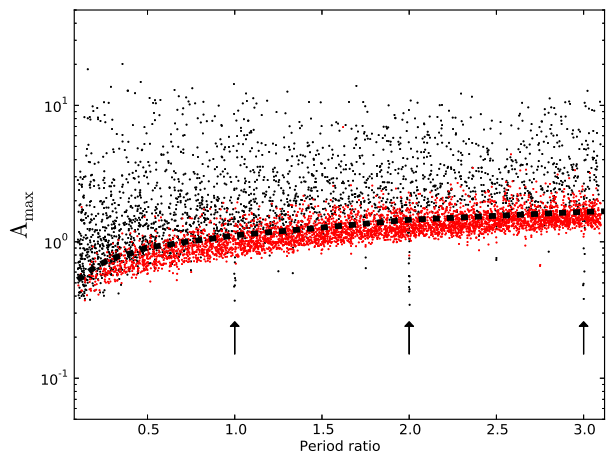}
\caption{The amplitude of the TTV measurements over the period ratio for
all individual simulations.
The red points indicate all simulations for which the determined period
has a FAP~$<10^{-3}$, which are the simulations with high TSNRs 
(see Fig.~\ref{fapsnr}). The black dashed line demonstrates the maximum
amplitude values given by Fig.~\ref{cirvsel}. The arrows indicate the 
localized drops of amplitudes at integer period ratios (see 
Sect.~\ref{Sec:SCT-res}).}
\label{amplrat}
\end{figure}

Figure~\ref{amplrat} shows $A_\mathrm{max}$ as a function 
of the period ratio for all individual simulations.
We divided the results in two groups:
Amplitudes coming from simulations with a FAP~$< 10^{-3}$ (red points) and
amplitudes with larger FAPs (black points).
For the latter group the uncertainties of 
the transit time measurements are high (TSNR is usually small)
and the amplitude primarily reflects the scatter of the TTVs due to noise.
For the first group the systematic variation due 
to spots is larger than the uncertainties (high TSNR)
and the amplitude represents the effect the 
spots have on the measurement of the TTVs.
The FAP~$< 10^{-3}$ points show much smaller 
scattering and are limited to a region close to the lower
edge of the measured amplitudes. As expected, the
majority of those points lie under the maximum deviation values (dashed line) which we
present in Fig.~\ref{amplrat}.
This region is basically defined by the planetary period and therefore the transit
duration, because the maximum amplitude of a spot-crossing 
event becomes smaller for smaller transit durations.

An interesting feature of Fig.~\ref{amplrat} are the localized minima of 
amplitudes at integer period ratios.
At these resonance periods the spot-crossing feature does hardly change
position in the transit profile and the spot-induced timing variation are small.
A period ratio of exactly one (two, three, etc.) 
means the TTVs are all on one level and vary only by noise.

\begin{figure}[bp]
\includegraphics[width=\linewidth]{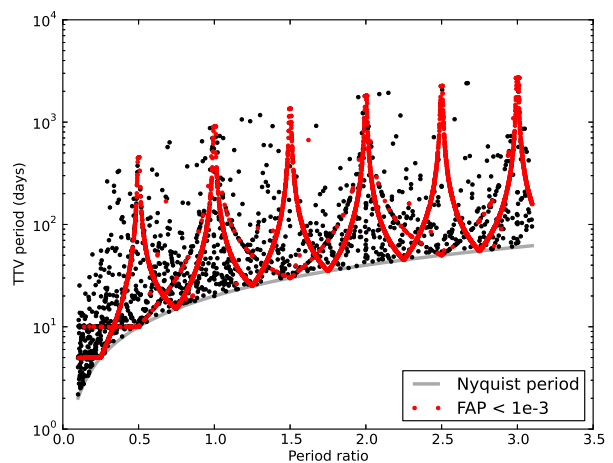}
\caption{The most-significant period determined from the TTVs over the input period ratio
for each individual simulation.
The gray line indicates the lower limit of possible periods defined by the Nyquist frequency.
The red points indicate periods with FAP~$< 10^{-3}$.}
\label{perrat}
\end{figure}

In Fig.~\ref{perrat} the period with the highest FAP 
determined from the $O-C$ diagram of
each individual simulation is plotted over the period ratio.
The lower envelope is determined by the Nyquist frequency.
The highest periods that can be determined are roughly defined by the time span covered by the TTVs.
The up-and-down patterns shown in Fig.~\ref{perrat} result from the different TSNRs of the simulations.
For low FAPs (red points) maxima at resonances of period ratios (0.5, 1, 1.5, etc.) and minima in between can be seen.
In a system where the periods of star and planet are not integral multiples,
the position of the spot-crossing event changes from one transit to the next. 
As the period ratio $P\sbsc{pl}/P_{\star}$ moves closer to a resonance, 
the difference between the longitudes $\lambda_\mathrm{o}$
of the spot-crossing events, in two consecutive transits, becomes smaller. 
This leads to a lower frequency TTVs, which is interpreted as higher TTVs
period by the periodogram.
Exactly at the resonance the TTVs are flat and the periods go down to the Nyquist frequency.
This is also true for period ratios at 0.5, 1.5, and so on;
however, here not every transit shows a deformation by a spot.
At ratio 0.5, for example, the spot is seen only every second transit,
which results in one level of transits that are not disturbed by the spot and a second level
of transits which are all shifted by the same value.
For very small noise the amplitude is then basically given by the difference between the two levels.

Going to larger FAPs, we see a second pattern (made up of red and black points in Fig.~\ref{perrat})
which is reminiscent of the previous structure but peaks only at integer resonances
and goes down to the Nyquist frequency at 0.5, 1.5, and so on.
This happens when the uncertainties of the individual mid-transit time measurements
are about equal to the amplitude caused by the spot-crossing effect.
For simulations with even higher FAPs, we end up with the points that are randomly distributed
because the uncertainties of the mid-transit times are larger
than potential spot-induced variations.

\subsection{Can spots mimic bodies in transit sequences?}

We emphasize again that our simulations show a worst case scenario.
In the majority of planetary systems,
periodic signals in TTVs caused by spots should be much weaker and less coherent.
In reality, the spots grow and decay, change their positions,
and probably have smaller sizes than what we assume in our simulations.
Such changes will always lead away from the clear patterns of
Fig.~\ref{perrat} (red points) to a random distribution of points.

However, we argue that it is possible to observe a periodic signal
mimicking TTVs of a potential companion in the planetary system caused by spot-crossing events.
For high TSNRs, period ratios close to resonances are particularly dangerous.
A careful inspection of the transit shapes and their deformations
should provide additional information on whether the determined mid-transit
times are reliable or subject to dispute.
In many cases the TTVs due to spots should show an odd behavior
as, e.g., a splitting of points into two or more levels,
or alternating intervals of strong and weak variability,
which could be valuable indications that the TTVs are possibly due to spots.
Unfortunately, depending on the exact properties of star, spot, and planet,
TTVs do not necessarily have to show such an odd behavior.

Whenever a significant periodic signal is found in noisy TTVs,
one should be careful with the interpretation,
especially if the star shows signs of activity.
If small scale features in the transit profiles are hard to discern anyway,
and the errors on the transit times are in the order of (or even larger than)
the deviations due to potential spots, one could still end up with a significant period.
Whether this period is caused by spots or a yet unknown body in the system will be hard to discriminate.

\begin{figure}[bp]
\includegraphics[width=\linewidth]{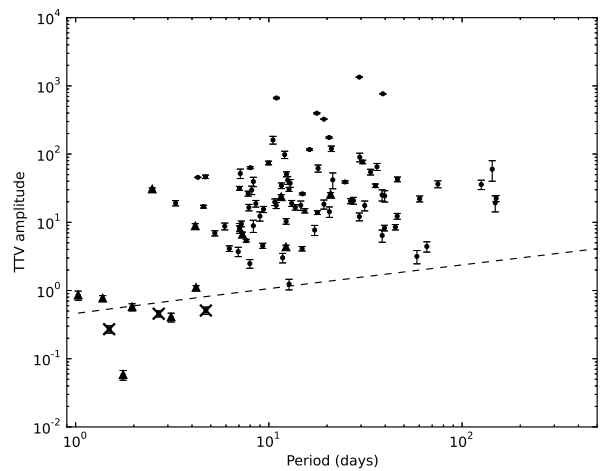}
\caption{
TTV amplitudes reported by \citetads{2013ApJS..208...16M}.
The dashed line denotes a linear fit to the
maximum amplitudes for TSNR=30 (see Fig.~\ref{cirvsel}).
The points marked with X are found to have periods related to the stellar rotation;
all of them lie well below our predicted amplitude maximum caused by spots.
The points marked with triangles are commented as spurious which could have been caused by
other sources capable of inducing TTVs (i.e. sampling or eclipsing binaries).}
\label{comppl}
\end{figure}

\section{Comparison to observations}
In Fig.~\ref{comppl} we show the cases of TTV detections as presented by \citetads{2013ApJS..208...16M}.
In their discussion regarding those TTVs detections \citetads{2013ApJS..208...16M}
suggest that three of them show a correlation
between the period of the TTVs and the rotational period of the star which was
determined from its light curve.
The TTV amplitudes for these planets
lie well below the upper limit we calculated from our simulations,
which supports our result that larger amplitudes caused by spots are unlikely.
Larger TTV amplitudes might be additionally modified by starspots
but are hardly exclusively created by them.

In a recent article \citetads{2015ApJ...806L..37M} present the light curve of the
KIC~6543674 triple star system, which consists of a short-period ($2.4$~days)
inner binary with a companion in a wide eccentric orbit.
Their study of the system presents Eclipse Timing Variations (ETVs)
with amplitudes of $\sim 4$~mins which do not seem to be caused by any known companion.
They imply these ETVs can be explained by the existence of starspots;
however, they do not model them.
Although we do not study ETVs of stars in this paper,
a spot-crossing event of a star should be comparable to our simulations with planets.
Spot-eclipses should cause shifts from the expected mid-eclipse time due to the deformation
of the eclipse shape as well.

The residual ETVs resemble quite closely the longitude-dependent spot-crossing pattern
we show, e.g. in Fig.~\ref{infvssnr}.
In order to create such ETVs, the period ratio between the orbital period of the occulting body
and the rotational period of the occulted star must be very close, but not exactly 
equal, to an integer value.
Interestingly, the ETVs do not show flat regions with no modulation,
which one would expect if no spot is on the visible hemisphere.
Instead the pattern continues very regularly, which indicates the existence
of two or more large active regions, non-uniformly distributed on the
surface of the occulted star.
Another interesting feature is the long lifetime of those active regions.
The modulation in the ETVs evolves so slowly that a reminiscent pattern repeats
for about three years before it significantly changes.

The amplitude of the measured ETVs appears to be 
larger than what our simulations predict for exoplanetary transits
with equal duration times;
the eclipses of the inner KIC~6543674 system last about $360$~min,
which according to Fig.~\ref{cirvsel} should lead to maximum amplitudes of about $2$~mins.
To check if we could reach the amplitude of the \citetads{2015ApJ...806L..37M} ETVs,
we run a simulation setup with the properties of this system and
obtain ETVs with comparable amplitudes in quite a few cases.
This result also agrees with the amplitudes of spot-induced ETVs from other binaries shown by \citetads{2015arXiv150307295O}.
The reasons for the deviation between the maximum amplitudes of spot-induced TTVs and ETVs are two:
(1) The shape of the eclipse and (2) the area of spots covered by the occulting body.
The eclipses of binary stars with comparable sizes usually show only the ingress and the egress
and not the flat part in between.
Therefore, spot-crossing anomalies can cause significant TTVs for spots at all possible longitudinal positions.
Due to the large size of the occulting body, it is also likely that large areas of
spots on the occulted body are eclipsed, which also leads to stronger ETVs.

It requires a detailed modelling of this system, including spots, to evaluate
whether these two arguments are sufficient to explain the observed ETVs -
or even if spots are an appropriate explanation for these ETVs at all.
Although this is beyond the scope of this work on exoplanets,
we emphasize that if these ETVs are really due to starspots they resemble the results of
our simulations quite nicely and demonstrate that clear repetitive modulations due to spots
over timescales of years are possible,
and the chances to misinterpret them as signals of an unknown body are not negligible.

\section{Discussion}
\label{Sec:Discussion}

Our simulations of spot-induced transit timing variations suggest
that even large spot-crossing anomalies in transit profiles should not lead to
TTV amplitudes larger than a few minutes (Figs.~\ref{cirvsel} and~\ref{amplrat}).
The amplitude depends strongest on the transit duration
and the position of the spot-crossing event.
As shown in Figs.~\ref{infvssnr}, the uncertainties of
the measured mid-transit times primarily depend on the TSNR and transit duration,
but only little on the characteristics of the spot-crossing events.
Specifically, the uncertainties do not depend on 
the position of the spot at the time of the transit,
whereas the TTV amplitude clearly does. Thus the uncertainties should not 
be used as a diagnostic for spot-induced TTVs, despite the deformation of 
the transit light curve, even for the rather large spots assuming in our simulations.
 
The amplitude of the spot induced TTVs is independent from the inclination of 
the star, as we discuss in Sect.\ref{Sec:res-starchar}. 
Nevertheless, in the case of low stellar inclination, a spot-crossing event is 
possible to occur at the cross section of the planetary orbit and the 
active latitudes of the star.
This has an impact on the results of Sect.~\ref{Sec:LCsims};
due to the fixed position of the spot-crossing events, 
the values of the TTVs amplitudes are going to be distinct and not continuous. 
Such $O-C$ diagrams produce low significance periodicities which have a low 
probability of being considered as real.

The maximum mid-transit time deviations derived from 
our simulations are significantly smaller than the maximum $O-C$ 
values calculated from \citetads{2015ApJ...800..142M} and \citetads{2013A&A...556A..19O}.
The reason for this discrepancy is our assumption that the in-transit
flux increment due to the spot occultation is not larger than $50$~\% of the mid-transit depth.
In our opinion this value is a better representation of observational evidence,
since we do not know of a single case were the induced anomaly is $\gtrsim 50$~\% of the transit depth.

\section{Conclusions}
\label{Sec:Conclusions}
Our simulations were geared to study TTVs induced by starspots
using high-precision photometric data without any red noise sources.
While it is relatively safe to apply our results to data obtained by space missions,
it is difficult to estimate the behavior of such spot-induced TTVs on light curves with
significant contribution of red noise (i.e., scintillation).

Finally, we summarize the most important results of our simulations
on spot-induced TTVs of exoplanets:
\begin{itemize}
 \item In agreement with the results of \citetads{2013A&A...556A..19O} and
       \citetads{2013MNRAS.430.3032B}, spot-crossing events can cause shifts of
       the measured mid-transit times.
       These shifts can cause a specific pattern depending on the longitude $\lambda\sbsc{o}$
       of the spot during the crossing event (see Fig.~\ref{infvssnr}).
 \item The amplitude of the shift is determined by the temperature of the spot relative to
       the photospheric temperature, the fraction of the planet covering spots,
       and the duration of the transit.
 \item The shape of the spot or the stellar inclination during a spot-crossing event
       do not seem to significantly affect the mid-transit times.
 \item We find only a weak dependence of the measurement uncertainties on longitude $\lambda\sbsc{o}$.
       We conclude that this is no reliable indicator for how much 
       individual timing measurements are affected by spots.
 \item The impact parameter of the planet and the the limb darkening 
       of the star influence the shape of the transit profile,
       which affects the TTV measurements. However, this seems to be a minor 
       contribution compared to the other influences. 
 \item A sequence of transits affected significantly by spots can  
       create a statistically significant periodical signal.
       In special cases the spot-induced TTVs could even show a clear, persistent modulation
       imitating the existence of a sub-stellar companion.
       However, these cases should be rare and a careful analysis of the data should indicate a potential
       influence of starspots, especially for systems with high TSNR.
 \item One should pay extra attention to TTVs with 
       amplitudes $\lesssim$1\% of the transit duration
       (see Figs.~\ref{cirvsel},~\ref{amplrat} and ~\ref{comppl}).
 \item The detection of significant spot-induced periods in TTVs depends on the TSNR,
       and according to our simulations is unlikely for TSNR~$\lesssim 15$.
       A detection of periods for TSNR~$> 15$ should be treated cautiously,
       especially if they are attributed to unknown bodies in the system.
\end{itemize}

\section{Acknowledgments}

PI acknowledges funding through the DFG grant RTG
1351/2 ''Extrasolar planets and their host stars''.
KFH is supported by DFG grant HU 2177/1-1.

\bibliographystyle{aa}
\bibliography{references}

\newpage

\end{document}